# VST TELESCOPE DYNAMIC ANALISYS AND POSITION CONTROL ALGORITHMS

P. Schipani, D. Mancini, Osservatorio Astronomico di Capodimonte, Napoli, Italy

Abstract

The VST (VLT Survey Telescope) is a 2.6 m class Alt-Az telescope to be installed on Cerro Paranal in the Atacama desert, Northern Chile, in the European Southern Observatory (ESO) site. The VST is a wide-field imaging facility planned to supply databases for the ESO Very Large Telescope (VLT) science and carry out stand-alone observations in the UltraViolet to Infrared spectral range. So far no telescope has been dedicated entirely to surveys; the VST will be the first survey telescope to start the operation, as a powerful survey facility for the VLT observatory. This paper will focus on the axes motion control system. The dynamic model of the telescope will be analyzed, as well as the effect of the wind disturbance on the telescope performance. Some algorithms for the telescope position control will be briefly discussed.

## 1 VST TELESCOPE DYNAMIC MODEL

The telescope coupling dynamic is so slow that it is possible to study independently the two main axis behaviors. The coupling of the two axes is usually negligible in a telescope, especially in the most important operating condition, i.e. the tracking phase, in which the axes trajectory is most of the time regular and not interested by strong accelerations.

Both axes structures are modeled by a number of inertias joined by stiffnesses and structural dampings (Figure 1 shows the altitude axis simplified model, the meaning of the symbols is reported in table 1). The axis gear is represented by the motor and teeth contact stiffness and by the damping. The inertia of the motor itself is taken into account, properly scaled by the transmission ratio. The structural data are derived from a Finite Element Analysis of the mechanical structure of the telescope.

The dynamic of the mechanical system can be described by second order differential equations in matrix form as:

$$J\ddot{\Theta} + F\dot{\Theta} + K\Theta = T$$

where J, F, K, T are the inertia, viscous damping, stiffness and torque matrix respectively, and $\Theta$ is the angular position vector. Figure 3 shows the open loop response of the altitude axis electromechanical model. Figure 4 shows the open loop transfer function of the azimuth axis. The first notch in the bode gain plots represents the Locked Rotor eigenfrequency (~10 Hz for both axes).

Table 1: Altitude axis structural parameters

| Parameter | Symbol |
|---|---|
| Center Piece inertia [kg·m$^2$] | J2a |
| M1 (Primary Mirror) inertia [kg·m$^2$] | J2b |
| Top Ring Inertia [kg·m$^2$] | J3a |
| M2 (Secondary Mirror) box Inertia [kg·m$^2$] | J3b |
| M2 Inertia [kg·m$^2$] | J3c |
| Motors inertia [kg·m$^2$] | J1 |
| Motors viscous friction [Nm/(rad/s)] | F1 |
| Viscous friction [Nm/(rad/s)] | F2 |
| Transmission damping [Nm/(rad/s)] | F12 |
| M1 pad damping [Nm/(rad/s)] | F2ab |
| Structural damping [Nm/(rad/s)] | F23 |
| Top Ring - M2 box damping [Nm/(rad/s)] | F3ab |
| M2 box - Mirror damping [Nm/(rad/s)] | F3bc |
| Transmission stiffness [Nm/rad] | K12 |
| M1 pad stiffness [Nm/rad] | K2ab |
| Structural stiffness [Nm/rad] | K23 |
| Top Ring - M2 box stiffness [Nm/rad] | K3ab |
| M2 box - Mirror stiffness [Nm/rad] | K3bc |
| Transmission ratio | R |

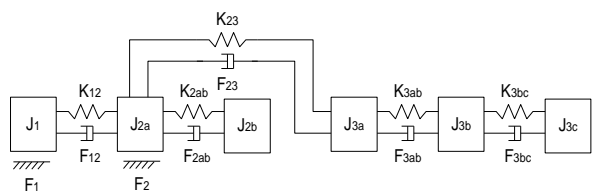

Figure 1: Altitude axis model

## 2 REQUIREMENTS FOR AXES CONTROL

Both VST azimuth an altitude axes are controlled in a double (position and speed) feedback control loop. The speed and position controllers have to be tuned to guarantee the two most important requirements:

- an extremely low tracking error
- a good disturbance rejection.

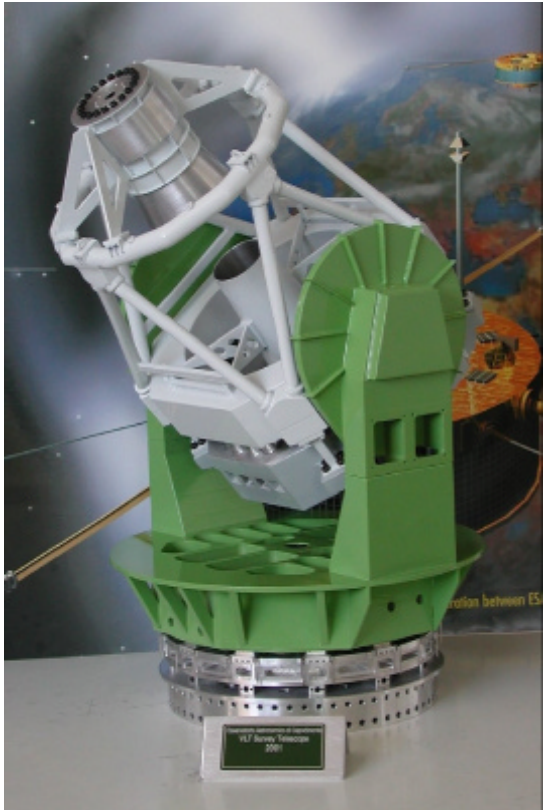

Figure 2: VST model

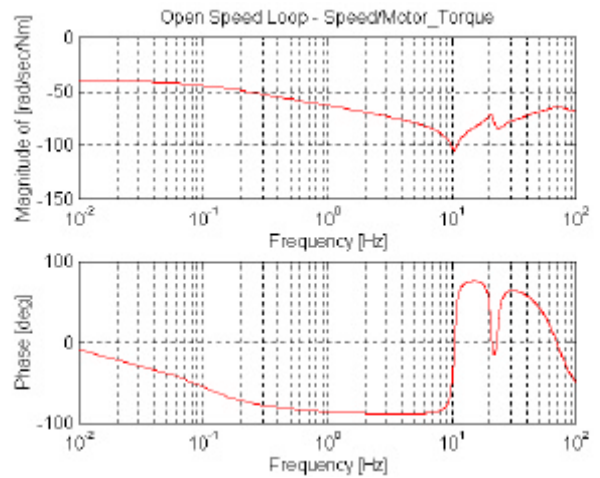

Figure 3: Altitude axis open loop transfer function bode diagram

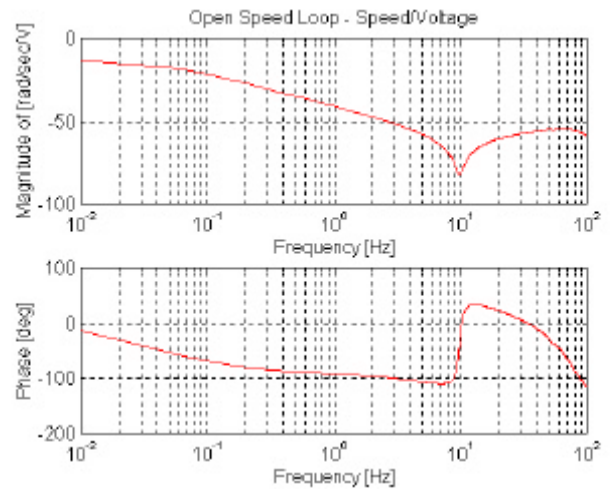

Figure 4: Azimuth axis open loop transfer function bode diagram

The absolute RMS tracking error must be as low as possible in order to guarantee the image quality during the observations. Furthermore, the stability robustness of the closed loop must be guaranteed with proper gain and phase margins, to reject structural parameter changes of the controlled system and/or environment modifications.

Particular attention must be paid in the synthesis of the controllers, in order to improve the guide and to assure a proper rejection against the disturbances. The main external disturbance to consider in this analysis is the wind shake. The wind mainly affects the altitude performance, because the altitude axis is exposed to a greater wind torque. The azimuth rotation is much more protected by the co-rotating enclosure. Actually most of the wind disturbance effect comes from the open slit of the enclosure in the observation direction, which affects mainly the altitude axis, and usually the influence of the wind disturbance on azimuth is neglected. The site chosen for VST, Cerro Paranal in the Atacama desert, is sometimes windy. Therefore a wind effect analysis has been carried out taking into account the Chilean site weather statistics.

## 3   CONTROLLERS

In this analysis the speed loop controller is a PI. The complexity of the real telescope controller could be increased by adding a series of Notch $2^{nd}$ order filters in order to attenuate eventual resonant spectral components.

The position controller chosen for this simulation is a standard Proportional+Integral (PI) controller, which ideally guarantees zero error to a ramp input (similar to the usual real observation conditions) after the transient phase. The real controller parameters will not be constant; pointing and tracking phases need different values for the proportional and integral constants. The problem can be circumvented or with a rough switch between different controller structure for small, medium and large errors, or providing the PI controller with anti-windup capability, or better with a variable structure controller (Scali, *et al.*, 1993) in which the parameters $K_p(e)$, $K_i(e)$ depend on the instantaneous position error value. This last solution has been proven to be very effective with other telescopes with similar dynamics (Mancini, *et al.*, 1997; Mancini, *et al.*, 1998). A detailed analysis of the dependence of the position controller on the tracking error is beyond the scope of the present simulation, which is limited to the "small" error case, i.e. to the tracking phase control. After the tuning of the position control parameters for both altitude and azimuth a -3db bandwidth of about 3 Hz has been obtained.

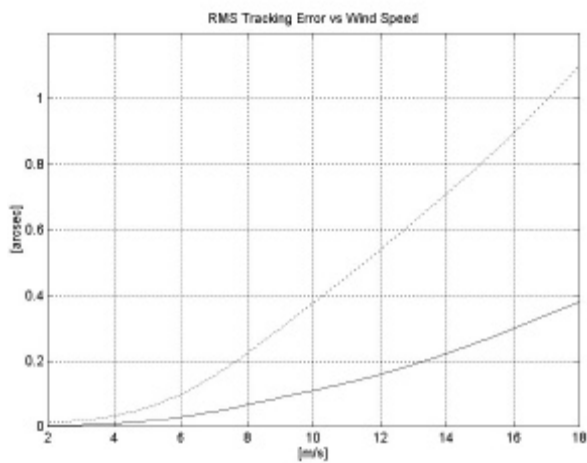

Figure 5: Expected errors vs wind speed, with wind speed reduction factors $\alpha=0.98$ (upper dotted line) and $\alpha=0.63$ (lower continuous line)

## 4   WIND DISTURBANCE EFFECT

The wind disturbance effect on the telescope performance has been studied using the Von Karman spectrum to model the wind speed power spectral density. Some simulations have been carried out at different wind speeds up to 18 m/s, the maximum operational speed in Paranal, in two case studies:
- considering a good wind speed reduction factor inside the enclosure, thanks to a proper setting of the wind-screens, and not observing in the wind direction ($\alpha=0.63$)
- in a pessimistic condition with almost no wind speed reduction ($\alpha=0.98$)

According to this simulation results the disturbance effect would increase with wind speed as foreseen; in the more realistic ($\alpha=0.63$) case, up to 12 m/s the effect could be considered not really performance limiting even in good seeing conditions, while at higher wind speeds in very good seeing conditions a negative effect could be noticed. Wind problems should be limited to a low percentage of telescope usage time; the wind speed is above 12 m/s in about 11% of the time, above 15 m/s only in the 5% of the time.

Table 2: RMS tracking errors due to wind disturbance

| V [m/s] | Time % With Wind Speed > V | RMS Tracking Error ($\alpha=0.98$) [arcsec] | RMS Tracking Error ($\alpha=0.63$) [arcsec] |
|---|---|---|---|
| 3 | 77 | 0.02 | 0.005 |
| 6 | 50 | 0.10 | 0.03 |
| 9 | 24 | 0.30 | 0.09 |
| 12 | 11 | 0.54 | 0.16 |
| 15 | 5 | 0.80 | 0.26 |
| 18 | 2 | 1.10 | 0.38 |